\documentclass[a4paper,fleqn,usenatbib,useAMS]{mnras}

\usepackage{graphicx}	
\usepackage{amsmath}	
\usepackage{amssymb,verbatim}	
\usepackage{multicol}        
\usepackage{bm}		
\usepackage{pdflscape}	

\usepackage[T1]{fontenc}
\usepackage{ae,aecompl}

\usepackage{newtxtext,newtxmath}

\title[PTA Constraints on Fermi Blazars]{Pulsar Timing Constraints on the Fermi Massive Black-Hole Binary Blazar Population}

\author[A. M. Holgado]{A. Miguel Holgado,$^{1}$\thanks{Contact e-mail: \href{mailto:holgado2@illinois.edu}{holgado2@illinois.edu}} 
Alberto Sesana,$^{2}$ 
Angela Sandrinelli,$^{3,4}$
Stefano Covino,$^{4}$ \newauthor
Aldo Treves,$^{3,4}$
Xin Liu,$^{1}$ and
Paul Ricker$^{1}$
\\
$^{1}$Department of Astronomy and National Center for Supercomputing Applications, University of Illinois at Urbana-Champaign, Urbana, IL 61801, USA \\
$^{2}$School of Physics and Astronomy and Institute for Gravitational-Wave Astronomy, University of Birmingham, Birmingham, B15 2TT, UK \\
$^{3}$Universit\'{a} dell'Insubria, Dipartimento di Scienza ed Alta Technologia, Via Valleggio 11, I-22100, Como, Italy \\
$^{4}$INAF-Istituto Nazionale di Astrofisica, Osservatorio Astronomico di Brera, Via Bianchi 46, I-23807 Merate (LC), Italy
}

\date{Last updated xx; in original form xx}

\pubyear{2018}

\graphicspath{{./figures/}}

\begin{document}
\label{firstpage}
\pagerange{\pageref{firstpage}--\pageref{lastpage}}
\maketitle

\begin{abstract}
  Blazars are a sub-population of quasars whose jets are nearly aligned with the line-of-sight, which tend to exhibit multi-wavelength variability on a variety of timescales. Quasi-periodic variability on year-like timescales has been detected in a number of bright sources, and has been connected to the orbital motion of a putative massive black hole binary.  
If this were indeed the case, those blazar binaries would contribute to the nanohertz gravitational-wave stochastic background.  
We test the binary hypothesis for the blazar population observed by the \textit{Fermi} Gamma-Ray Space Telescope, which consists of BL Lacertae objects and flat-spectrum radio quasars.
Using mock populations informed by the luminosity functions for BL Lacertae objects and flat-spectrum radio quasars with redshifts $z \le 2$, we calculate the expected gravitational wave background and compare it to recent pulsar timing array upper limits.
The two are consistent only if a fraction $\lesssim 10^{-3}$ of blazars hosts a binary with orbital periods $<5$ years. We therefore conclude that binarity cannot significantly explain year-like quasi-periodicity in blazars.

\end{abstract}




\begingroup
\let\clearpage\relax
\endgroup
\newpage

\section{Introduction} \label{sec:intro}
Current observational searches for supermassive black-hole (SMBH) binaries on a variety of scales -- from sub-pc to kpc separations -- and with masses ranging within  $10^6 M_\odot - 10^9 M_\odot$ are motivated by a better understanding of both galaxy formation and gravitational-wave (GW) astrophysics.
There are several methods to characterize individual SMBH binaries that are thought to reside in some active galactic nuclei (AGN), including direct imaging \citep[e.g.,][]{rodriguez_compact_2006,bansal_constraining_2017}, long-term photometric monitoring \citep[e.g.,][]{graham_systematic_2015,charisi_population_2016,liu_systematic_2016,sandrinelli_gamma-ray_2017}, and multi-epoch spectroscopy of AGN broad lines \citep{shen_constraining_2013,liu_constraining_2014,runnoe_large_2015,runnoe_large_2017,wang_searching_2017}.  
Sub-pc SMBH binaries are of particular interest since they are in the regime where GW emission starts to drive their inspiral and occurs at frequencies where pulsar timing arrays \citep[PTAs,][]{1990ApJ...361..300F} are directly sensitive to.  
In this Letter, we focus on the blazar population and its contribution to the nanohertz GW stochastic background assuming a fraction of the population is composed by SMBH binaries. 
\par
Blazars are a particular type of AGN in which the jet is almost aligned with the line-of-sight (i.e., $\lesssim 5^\circ$) as described by the AGN unification model \citep[e.g.,][]{urry_unified_1995,urry_blazars_2000}. 
The {\it Fermi} Gamma-Ray Space Telescope has provided the most complete all-sky catalog of blazars out to redshift $z \lesssim 4$ \citep[e.g.,][]{acero_fermi_2015,ackermann_gamma-ray_2017}.  
Blazars may exhibit multi-wavelength quasi-periodicities on a variety of timescales which have been proposed to possibly range from days to years \citep[e.g.,][]{ackermann_multiwavelength_2015,sandrinelli_quasi-periodicities_2016,sandrinelli_quasi-periodicities_2018}.
Quasi-periodic light curves on timescales ranging from months to years has been interpreted by some to be due to binary orbital motion \citep[e.g.,][]{de_paolis_astrophysical_2002,rieger_supermassive_2007}.
The quasi-periodic features may also be due to intrinsic variability in the accretion disk such as low-frequency quasi-periodic oscillations \citep[e.g.,][]{king_quasi-periodic_2013}.  
Even though binary models have been used to explain the observations and tests for relativistic Doppler boosts have been performed \citep[e.g.,][]{dorazio_relativistic_2015,charisi_testing_2018,yan_testing_2018}, most candidates cannot definitively be confirmed nor ruled out with such tests alone.  
With the advent of GW astronomy, a GW signal or lack thereof can provide an additional constraint on such binaries and help to confirm or rule out the binary hypothesis as described below.  
\par
PTAs are expected to detect the stochastic nanohertz gravitational-wave background (GWB) from the inspiral of the SMBH binary population out to redshifts $z \lesssim 2$.
The three international collaborations that aim to detect GWs in this regime are the North American Nanohertz Observatory for Gravitational Waves  \citep[NANOGrav,]{demorest_limits_2013,arzoumanian_gravitational_2014}, the Parkes Pulsar Timing Array \citep[PPTA,][]{reardon_timing_2016}, and the European Pulsar Timing Array \citep[EPTA,][]{desvignes_high-precision_2016}, which together form the International Pulsar Timing Array \citep[IPTA,][]{verbiest_international_2016}.
\par
The binary hypothesis, i.e., the supposition that quasars host a SMBH binary if they exhibit certain photometric and/or spectroscopic features, has recently received attention in the context of GW astrophysics.  
\cite{sesana_testing_2018}, S18 herein, tested the binary hypothesis for SMBH binary candidates selected via photometric periodicity from the Catalina Real-time Transient Survey \citep[CRTS,][]{graham_systematic_2015} and the Palomar Transient Factory \citep[PTF,][]{charisi_population_2016}.  
S18 computed the implied SMBH binary merger rate from CRTS and PTF candidates and found the inferred GW background to be in moderate-to-severe tension with current PTA upper limits, depending on the corrections for incompleteness, selection effects, virial-mass distributions, and mass-ratio distributions.
There are also a number of binary blazar candidates that have been reported in the literature based on their quasi-periodic oscillations \citep[e.g.,][]{romero_beaming_2000,valtonen_massive_2008}.  
\cite{sandrinelli_quasi-periodicities_2014,sandrinelli_quasi-periodicities_2018} and \cite{ackermann_multiwavelength_2015} recently reported two binary blazar candidates, PKS 2155-304 and PG1553+113, which exhibit evidence for multi-wavelength quasi-periodic modulations.
The gamma-ray light curve in particular shows quasi-periodic behavior on year-like timescales.  
Follow-up studies have used binary models to explain the quasi-periodic light-curve features \citep{caproni_jet_2017,tavani_blazar_2018}.  
We caution that evidence for quasi-periodic features may not always be statistically significant.  
\par
In this Letter, we test the binary hypothesis for the blazar population, which consists of BL Lacertae objects (BL Lacs) and flat-spectrum radio quasars (FSRQs).    
\cite{ajello_luminosity_2012}, A12 herein, and \cite{ajello_cosmic_2014}, A14 herein, obtained blazar luminosity functions for FSRQs and BL Lacs, respectively.
We use these luminosity functions to estimate the cosmic merger rate of SMBH binaries, assuming a given binarity fraction.
The cosmic merger rate is then integrated to get the background strain amplitudes and compared to the most recent PTA upper limits.  
For our test of the binary hypothesis, we will consider the scenario in which as much as 10\% of blazars show quasi-periodicity that is statistically significant \citep{sandrinelli_quasi-periodicities_2018}, which we will assume to be due to binarity.
We adopt the same cosmology as A12 and A14 ($H_0 = 71 \ {\rm km} \, {\rm s}^{-1} \, {\rm Mpc}^{-1}$ and $\Omega_{\rm M} = 1-\Omega_{\Lambda} = 0.27$). 
\par
%
\section{The Cosmic Blazar Population and Gravitational-Wave Backgrounds}
%
\subsection{Blazar subpopulations}
\begin{figure}
\centering
\includegraphics[width=\columnwidth]{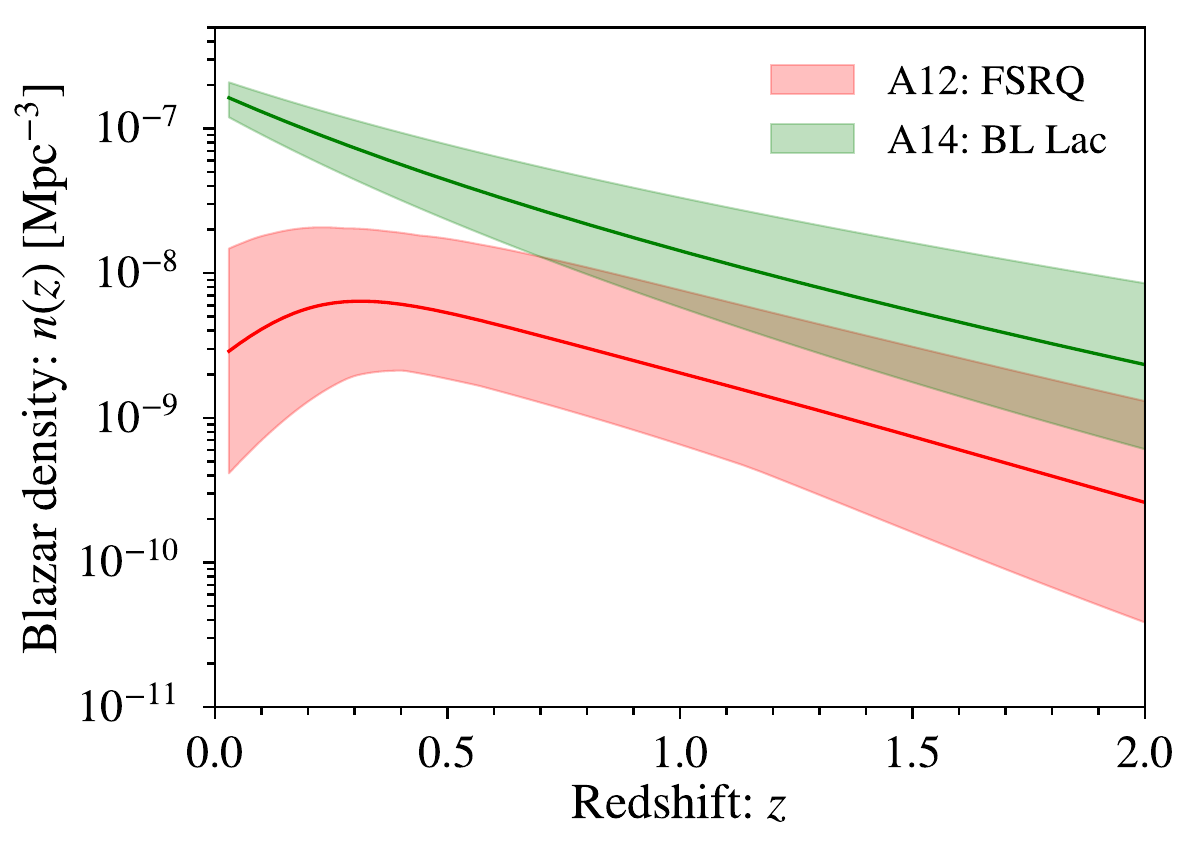}
\caption{\label{fig:density} Realizations of the luminosity-dependent density evolution model for FSRQs (A12) and BL Lacs (A14).  
The red and green lines correspond to the FSRQ and BL Lac models evaluated at the best-fit parameters reported in A12 and A14, respectively. 
The shaded regions represent the $1\sigma$ uncertainties in these models.  
}
\end{figure}

Blazars consist of two subpopulations: BL Lacs, which have a relatively featureless optical emission continuum, and FSRQs, which show optical emission lines.
A12 and A14 obtained models for the blazar luminosity functions for FSRQs and BL Lacs, respectively, which are defined as
\begin{equation}
\frac{{\rm d}^3 N}{{\rm d}L_\gamma \, {\rm d}z \, {\rm d}\Gamma} = 
\frac{{\rm d}^3 N}{{\rm d}L_\gamma \, {\rm d}V \, {\rm d}\Gamma} \frac{{\rm d}V}{{\rm d}z} = \Phi\left(L_\gamma, V(z), \Gamma\right) \frac{{\rm d}V}{{\rm d}z} \ ,
\end{equation}
where $L_\gamma$ is the gamma-ray luminosity, $\Gamma$ is the photon index, and ${\rm d}V/{\rm d}z$ is the differential comoving volume per differential redshift. 
We use the luminosity-dependent density evolution model for the observed blazar population of A12 and A14.
A12 and A14 also present beaming corrections to these observed luminosity function to obtain intrinsic luminosity functions that represent the parent population of blazars. 
We account for these beaming corrections when we estimate the cosmic merger rate as described in \S \ref{ssec:background}.
The main quantity of interest extracted from the luminosity function is the number density of blazars, which we obtain by integrating the luminosity function over all photon indices and luminosities:
\begin{equation}
n(z) = \frac{{\rm d}N}{{\rm d}V} = \int_{L_{\gamma, \rm min}}^{L_{\gamma, \rm max}} \int_{\Gamma_{\rm min}}^{\Gamma_{\rm max}} \Phi(L_\gamma, z, \Gamma) \ {\rm d}\Gamma \, {\rm d}L_\gamma \ .
\end{equation}
Integration boundaries are reported in A12 and A14.
We plot the number density for the BL Lacs and FSRQs in Figure \ref{fig:density}, considering only the population at redshifts $z\le 2$,  which is sufficient to account for the bulk of the GW signal to which PTAs are sensitive to \citep{2008MNRAS.390..192S}.
%
As A14 notes, the increase in the BL Lac number density at $z\lesssim 0.5$ is due to contributions from the high-synchrotron-peaked BL Lacs. 
A14 also notes that the turnover in the FSRQ number density at $z\lesssim 0.5$ may imply an evolutionary connection between BL Lacs and FSRQs \citep[e.g.,][]{bottcher_evolutionary_2002,ghisellini_transition_2011}.  
We do not consider such implications in this Letter and solely focus on the binary fraction of each population as independently contributing to the nanohertz GW background. 
\subsection{Blazar mass distributions}
\begin{figure}
\centering
\includegraphics[width=\columnwidth]{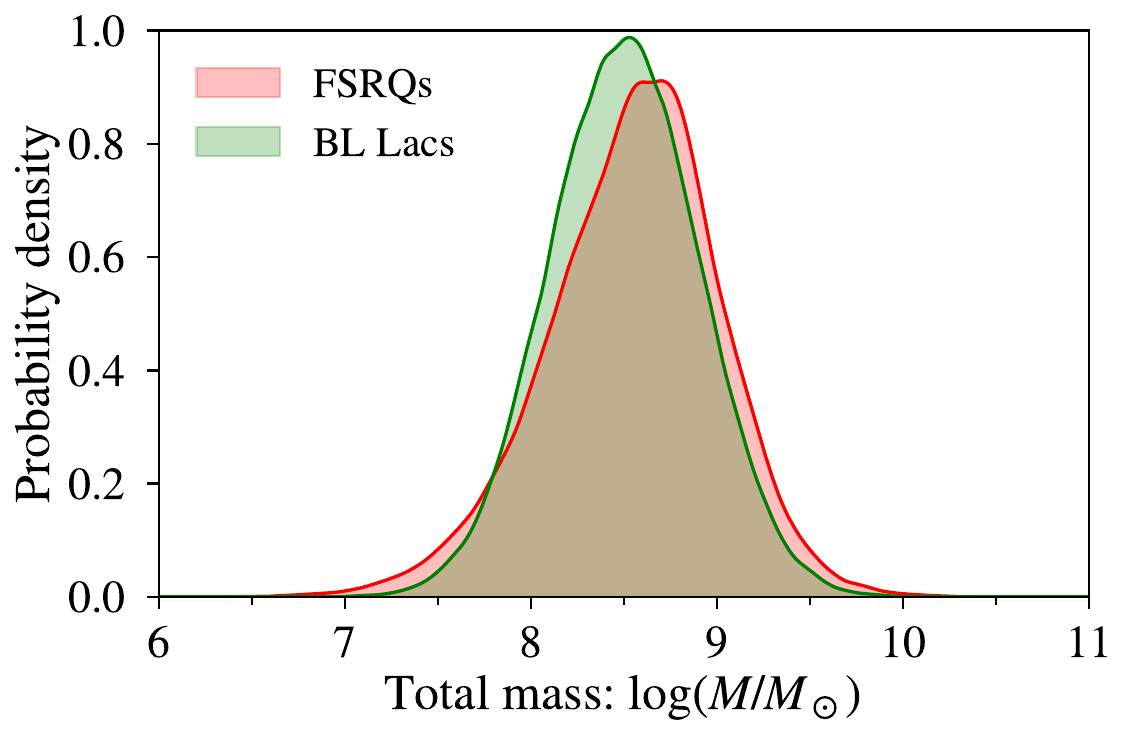}
\includegraphics[width=\columnwidth]{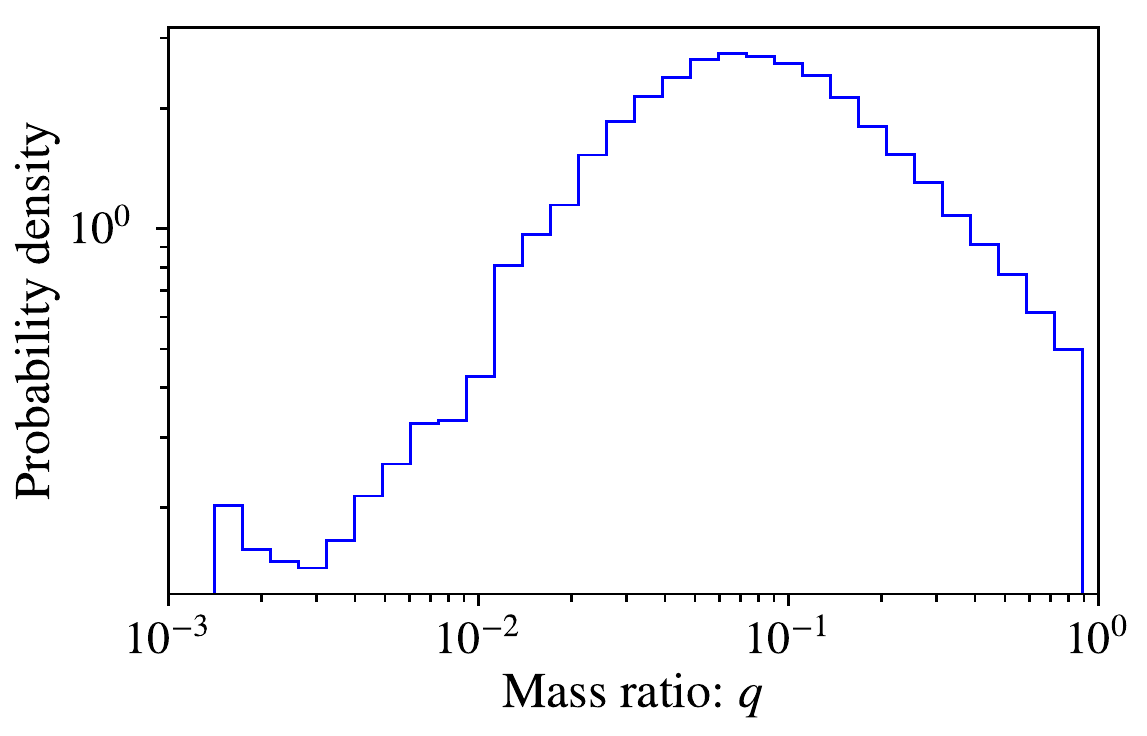}
\caption{\label{fig:mdist} Top panel: mass distribution of {\it Fermi} blazars.
BL Lac masses are often obtained via the $M-\sigma$ relation while FSRQ masses are often obtained with the virial method.
We have applied bias corrections to the FSRQ virial-mass distribution \citep{shen_biases_2008,sesana_testing_2018} to obtain an estimate of the true distribution of FSRQ masses.
Bottom panel: mass-ratio distribution for SMBH binaries from the Millennium simulation \citep[e.g.,][]{springel_simulations_2005}. 
}
\end{figure}

The BH mass is a particularly important parameter for our calculations because the strain amplitude scales with the chirp mass as $h_{\rm c} \propto {\cal M}^{5/3}$.
Since BL Lacs do not show emission lines in their spectra, host-galaxy relations -- like the $M-\sigma$ -- are generally used to infer the BH mass \citep[e.g.,][]{gebhardt_relationship_2000,ferrarese_fundamental_2000,barth_stellar_2002,falomo_black_2002,tremaine_slope_2002,barth_black_2003,plotkin_dynamical_2011,kormendy_coevolution_2013}.
The BH masses for FSRQs, instead, are typically obtained via virial methods \citep[e.g.,][]{castignani_black-hole_2013}.
We therefore apply the correction described in \cite{shen_biases_2008} and S18 to obtain estimates for the true FSRQ mass distribution from the measured FSRQ virial masses.  
We plot the mass distributions for the BL Lacs and FSRQs in the top panel of Figure \ref{fig:mdist}.  
The mass distributions for the FSRQ and BL Lac populations are fairly similar, which is expected given that there is no evidence in the features, variability, or inclination of these blazar subpopulations pointing to an intrinsic different nature of the accreting SMBH. 
\subsection{Generating blazar mock populations and computing the GW background} \label{ssec:background}
Following \cite{phinney_practical_2001}, the characteristic strain spectrum from a population of circular SMBH binaries undergoing GW-dominated inspiral is
\begin{equation} \label{eq:strain}
h_{\rm c}^2(f) = \frac{4G}{\pi c^2 f^2} \int_0^\infty {\rm d}z \int_0^\infty {\rm d}{\cal M} \, \frac{{\rm d}^2 n}{{\rm d}z \, {\rm d}{\cal M}} \frac{1}{1+z} \frac{{\rm d}E_{\rm GW}({\cal M})}{{\rm d}\ln f_{\rm r}} \ ,
\end{equation}
where $z$ is the redshift, ${\cal M}$ is the chirp mass and obeys ${\cal M} = M q^{3/5}/(1+q)^{6/5}$, with $M=M_1+M_2$ being the total mass and $q = M_1/M_2 \le 1$ as the mass ratio, ${\rm d}^2n/{\rm d}z{\rm d}{\cal M}$ is the cosmic merger rate, and the energy radiated per logarithmic frequency interval is
\begin{equation}
\frac{{\rm d}E}{{\rm d}\ln f_{\rm r}} = \frac{1}{3}G^{2/3}{\cal M}^{5/3} (\pi f_{\rm r})^{2/3} \ .
\end{equation}
We discuss in \S\ref{sec:results} why our assumption of a circular SMBH binary population (i.e., no eccentricity due to environmental coupling) does not significantly affect our estimates.
Given the A12 and A14 models for the complete blazar population, some fraction may be binaries, which contributes to the GW background.
We thus approximate the cosmic merger rate as
\begin{equation}
\frac{{\rm d}^2 n}{{\rm d}z \, {\rm d}{\cal M}} = \frac{{\rm d}^2 n}{ {\rm d}{\cal M} \, {\rm d}t_{\rm r}} \frac{{\rm d}t_{\rm r}}{{\rm d}z} 
\approx \frac{\eta}{\zeta\tau_{\rm co}}\frac{{\rm d}n}{{\rm d}{\cal M}} \frac{{\rm d}t_{\rm r}}{{\rm d}z} \ ,
\end{equation}
where $\eta$ is the average binary fraction, $\zeta$ is the beaming correction factor, and $\tau_{\rm co}$ is the coalescence timescale for a GW-driven, circular SMBH binary emitting at a rest-frame GW frequency $f_{\rm r}=2 f_{\rm orb}$, given by
\begin{equation}
\tau_{\rm co} ({\cal M},f_{\rm r}) = \frac{5}{256} \left(\frac{G{\cal M}}{c^3}\right)^{-5/3} (\pi f_{\rm r})^{-8/3} \ .
\end{equation}
To complete the calculation, we need to estimate the chirp-mass distribution, ${\cal P}({\cal M})$, and the coalescence time, $\tau_{\rm co}$.
The chirp-mass distribution is estimated from the FSRQ and BL Lac mass functions by assuming a SMBH binary mass-ratio distribution from the Millennium simulation \citep[e.g.,][]{springel_simulations_2005}, which is plotted in the in bottom panel of Figure \ref{fig:mdist}.
Cosmological hydrodynamical simulations like Illustris may predict a flatter mass-ratio distribution \citep{kelley_massive_2017}, though we discuss later in \S\ref{sec:results} that such differences in reasonable mass-ratio distributions do not have a significant effect on the GW background amplitude. 
\par
Drawing from the blazar luminosity functions, mass distributions, and mass-ratio distribution of the Millennium simulation, we generate $N=5{\times}10^5$ realizations for both blazar subpopulations.
For each value of $({\cal M},z)$, the coalescence time is then the one corresponding to the longest orbital period that {\it Fermi} probes, which we take to be $P_{\rm orb} = 5 \ {\rm yr}$.
The reasoning for why the longest orbital period that {\it Fermi} is sensitive to is used for the coalescence time for the sample population as opposed to each sample's respective orbital period is as follows.
Suppose that $N$ identical binaries (same in mass and redshift) with different periods $P_i < P_{\rm orb}$, are identified in the {\it Fermi} sample.
The merger rate of this sample binary population is
\begin{equation}
\dot{N} \equiv \frac{{\rm d}N}{{\rm d}t_{\rm r}} = \frac{N(P_i < P_{\rm orb})}{\tau_{\rm co}} \ , 
\end{equation}
where $\tau_{\rm co}$ is the coalescence time evaluated at the longest orbital period that {\it Fermi} is sensitive to. 
In other words, because of continuity, the merger rate is simply the number of observed binaries $N$ divided by the time each individual system would be observable as a binary by {\it Fermi}. 
Further details can be found in S18.  
For the average binary fraction, we take $\eta = 0.1$, which is motivated by the fraction of bright blazars that exhibit year-timescale quasi-periodic behavior \citep{sandrinelli_quasi-periodicities_2018}.
A12 reported a beaming correction factor of $\zeta_{\rm F} \approx 0.001$ for FSRQs, corresponding to their reported average jet tilt angle of $\sim 5^\circ$.
For BL Lacs, we use $\zeta_{\rm B} \approx 0.0025$, corresponding to A14's reported jet tilt angle of $\sim 10^\circ$.  
\par
The strain is usually expressed in terms of the GW amplitude at a nominal reference frequency of 1 yr$^{-1}$, as $h_{\rm c} = A_{\rm yr}\left(f / 1 \, {\rm yr}^{-1}\right)^{-2/3}$.
After obtaining the characteristic strain spectra with Eq.~\eqref{eq:strain}, we evaluate the amplitudes $A_{\rm yr}$, which we then compare to PTA upper limits.  
Given that PTAs are most sensitive to the most massive binaries at lower redshifts, we expect that the GW amplitude distributions we compute will be most sensitive to what the demographics of our mock populations are at lower redshifts.  
\subsection{Quantifying tension with PTA upper limits}
To compare the theoretical distributions to observations we use the same framework adopted by S18. We define the odds ratio for the null-hypothesis model over the binary-hypothesis model as
\begin{equation}
\Lambda_{\rm NB} = \frac{{\cal P}({\rm D}|{\rm N})}{{\cal P}({\rm D}|{\rm B})} \frac{{\cal P}({\rm N})}{{\cal P}({\rm B})} = \frac{{\cal P ({\rm D}|{\rm N})}}{{\cal P}({\rm D}|{\rm B})}\ ,
\end{equation}
where we have taken the null hypothesis and binary hypothesis to be {\it a priori} equally probable such that ${\cal P}({\rm N}) = {\cal P}({\rm B})$.  
The likelihood of model X is
\begin{equation}
{\cal P}({\rm D}|{\rm X}) = \int {\cal P}_{\rm PTA} (A) {\cal P}_{\rm X} (A) \ {\rm d}A \ ,
\end{equation}
where ${\cal P}_{\rm PTA}(A)$ is the posterior amplitude distribution returned by the analysis of PTA data. Following S18, we quantify tension with the PPTA upper limits \citep{shannon_gravitational_2015}, 
where we approximate the PPTA posterior amplitude distribution as a Fermi function of the form
\begin{equation}
{\cal P}_{\rm PTA} (A) = \frac{C_1}{e^{(A-C_2)/C_3} + 1} \,.
\end{equation}
where $C_1 = 1.63$, $C_2 = 1.2{\times}10^{-16}$, and $C_3 = 2.6{\times}10^{-16}$.
The probability of the null hypothesis is $p_{\rm N} = p({\rm D}|{\rm N})/(p({\rm D}|{\rm N})+p({\rm D}|{\rm B}))$ and the probability of the binary hypothesis is $p_{\rm B} = 1 - p_{\rm N}$.

\section{Results and Discussion} \label{sec:results}
\begin{figure}
\centering
\includegraphics[width=\columnwidth]{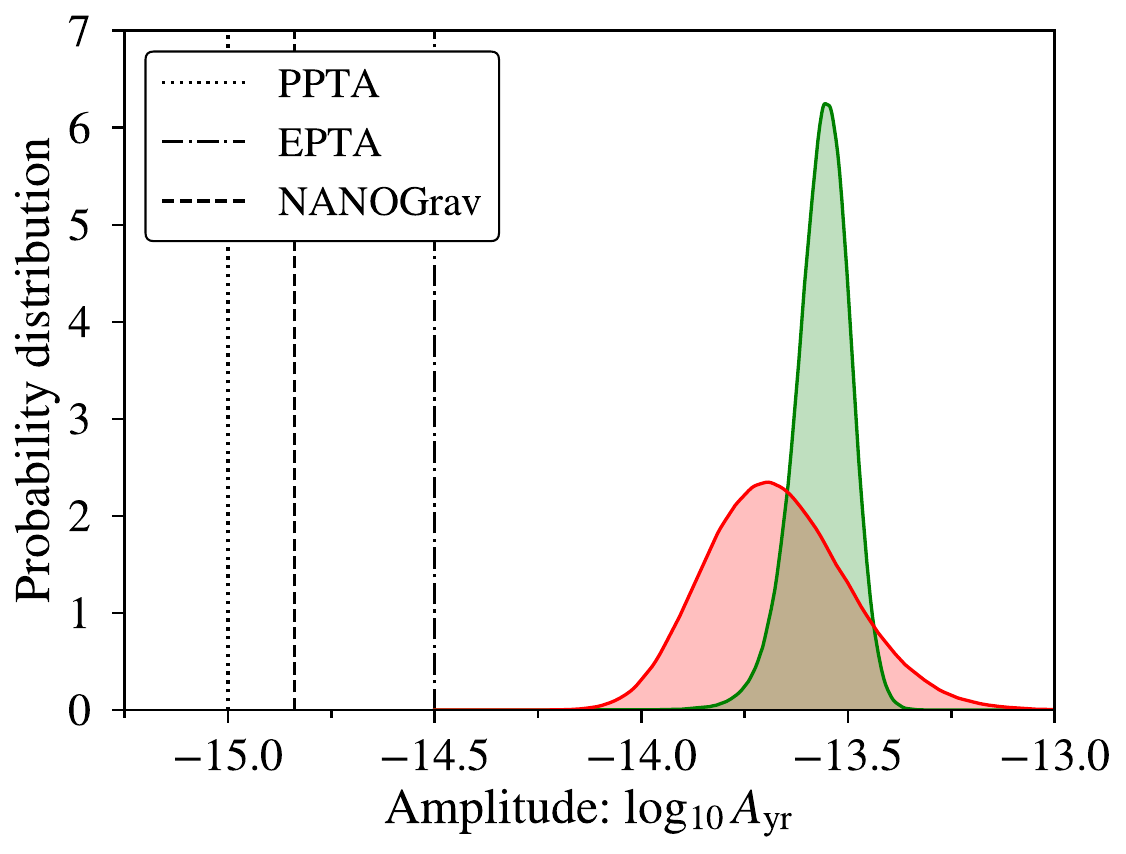}
\caption{\label{fig:adist} Amplitude distributions for the BL Lac and FSRQ populations assuming that 10\% of blazars are binaries \citep{sandrinelli_quasi-periodicities_2018}, i.e., $\eta = 0.1$, and with the beaming correction factors of $\zeta_{\rm B} = 0.0025$ (A14) and $\zeta_{\rm F} = 0.001$ (A12) for BL Lacs and FSRQs, respectively.  
The dotted, dashed, and dot-dashed vertical lines are 95\% upper limits on $A_{\rm yr}$ from PPTA, NANOGrav (11-yr), and EPTA, respectively. 
}
\end{figure}

\begin{figure}
\centering
\includegraphics[width=\columnwidth]{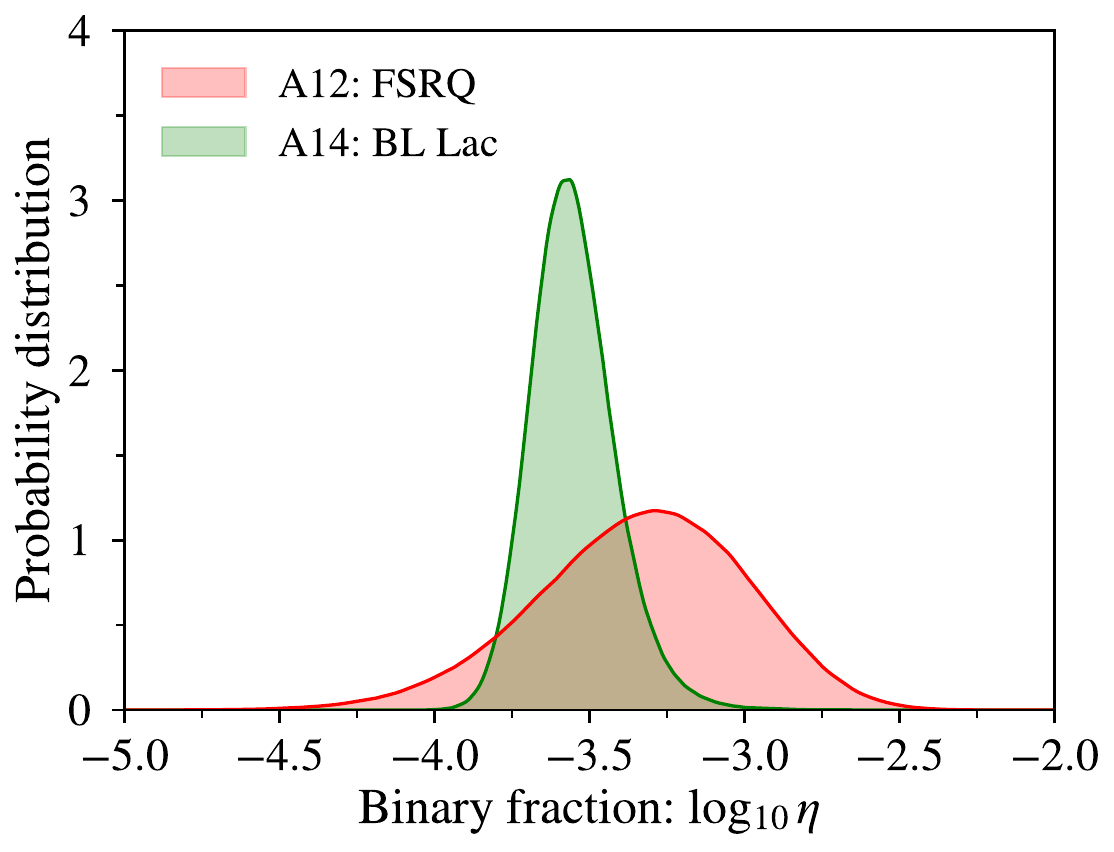}
\caption{\label{fig:binfrac} Distributions of the maximum binary fraction for the BL Lac and FSRQ populations allowed by the NANOGrav 11-yr 95\% upper limit of $A_{\rm yr} = 1.45{\times}10^{-15}$.     
}
\end{figure}

With amplitude distributions for the BL Lac and FSRQ populations in hand, we can proceed to the comparison with recent PTA upper limits.
We plot the amplitude distributions and the reported 95\% upper limits from the PPTA NANOGrav 11-yr, and EPTA in Figure \ref{fig:adist}.  
%
Both the BL Lac and FSRQ distributions show severe tension with all PTA measurements. We quantify this tension in Table \ref{tab:odds}, using the PPTA posterior distributions as test case. 
The odds ratios decisively favor the null hypotesis, and the probability that $\eta = 0.1$ of the BL Lacs and FSRQs are binaries is around $10^{-8}$. 
We can therefore ask what binarity fraction at periodicities probed by {\it Fermi} would be consistent with PTA data. 
Using the most recent NANOGrav 11-year upper limits (other limits yield quantitatively similar results), we obtain binary fraction upper limits $\eta \lesssim 10^{-3}$ for both populations, as shown in Figure \ref{fig:binfrac}.
Therefore, no more than a blazar in a thousand can host a SMBH binary with period $P<5$ yr.
%
%
%
\begin{table}
\centering
\caption{Model selection using posterior distributions from the PPTA between the null hypothesis and binary hypothesis for the FSRQ and BL Lac populations with their respective beaming corrections and assuming $\eta = 0.1$.
}
\label{tab:odds}
\begin{tabular}{ccc}
\hline \hline
Model Pair & $\log_{10} (\Lambda_{\rm NB})$ & $p_{\rm B}$ \\ \hline
N/FSRQ & 7.14 & $7.23{\times} 10^{-8}$ \\
N/BL-Lac & 8.31 & $4.81{\times}10^{-9}$ \\
\hline
\end{tabular}
\end{table}
%
\par
Although this calculation relies on a number of assumptions, they are unlikely to significantly alter this main result.
Despite our assumption of a specific mass-ratio distribution, S18 and \cite{inayoshi_gravitational_2018} have shown that reasonable mass-ratio distributions can affect the overall GW background amplitude by no more than a factor of $\approx 2$, thus leaving our main result effectively unchanged.
Similarly, the GW background amplitude has a strong dependence on the assumed maximum orbital period to which {\it Fermi} is sensitive to, which we conservatively assumed to be 5 years, thus making our results robust.
We also notice that PTAs are sensitive to sub-pc SMBH binaries regardless of whether they are active or quiescent.
Our calculation is representative of the true GW background only if SMBH binaries are assumed to form solely in gas-rich mergers, thus being harbored in quasars or blazars, i.e., in AGNs in general.
If conversely, SMBH binaries are not connected to AGN activity, our estimates need to be corrected for the active duty cycle of {\it Fermi} blazars.
Again, such correction goes in the direction of making the GW background larger, thus strengthening our results.
We also assumed that the observed periodicity is related to the binary {\it orbital} period, whereas binarity might induce periodicity through precession.
We note, however, that precession timescales are much longer than the binary orbital period.
An observed precession on year-like timescales would imply a much more compact binary, thus implying a much shorter coalescence time $\tau_{\rm co}$, and consequently a much higher GW background.
\par
There are also a few dynamical effects that can partially suppress the expected GW background in the PTA band.
In fact, we did not consider environmental coupling in this work since S18 finds that the GW signal amplitude would decrease only by a factor of ${\sim} 1.5$ at $6 \ {\rm nHz}$.
Moreover, their estimate does not account for the fact that increased eccentricity due to environmental coupling decreases the coalescence timescale for SMBH binaries, which increases the cosmic merger rate, and ultimately, the GW background strain amplitude. 
%
%
\section{Conclusions} \label{sec:conclude}
In this Letter, we have used PTA upper limits on the stochastic GW background at nanohertz frequencies to constrain the nature of quasi-periodic blazar candidates that {\it Fermi} observes. 
The {\it Fermi} blazar catalog contains the most complete all-sky census of $\gamma$-ray blazars out to redshifts of $z\lesssim 4$ and statistically significant quasi-periodicity on year-like timescales has been detected in some of the brightest sources.
One interpretation of this observed quasi-periodicity on year-like timescales is that it stems from the orbital motion of a SMBH binary.
Is this is indeed the case, then the collective population of periodic blazars would produce a stochastic GW background detectable in the PTA frequency band.
\par
As previously demonstrated by S18, we have shown that even though PTAs have not yet detected the nanohertz GW stochastic background, the upper limits that they place provide astrophysically relevant constraints for characterizing the SMBH binary population.
We have also presented upper limits on the binary fraction of BL Lacs and FSRQs of the blazar population, where we have used blazar luminosity functions, distributions for the measured masses from host-galaxy relations for BL Lacs and corrected virial masses for FSRQs, and mass-ratio distributions informed by the Millennium simulation.
We conclude that binarity alone cannot explain the quasi-periodic behavior of blazars.
\par
Further study of current binary candidates and future searches for sub-pc SMBH binaries will continue.
We expect the methods presented in S18 and this Letter to be pivotal tests to favor or disfavor SMBH binary candidates.  
\section*{Acknowledgements}
AMH is supported by the DOE NNSA Stockpile Stewardship Graduate Fellowship.
AMH gratefully acknowledges the NANOGrav PIRE program for supporting travel to the University of Birmingham, UK, where much of this work was carried out. 
AMH also thanks the School of Physics and Astronomy and the Institute for GW Astronomy for their hospitality during his visit to Birmingham.  
AS is supported by the Royal Society. 
We thank Maria Charisi, Eliu Huerta, Vicky Kalogera, Xavier Siemens, Joseph Lazio, Stephen Taylor, Chiara Mingarelli, Maura McLaughlin, and Zoltan Haiman for helpful comments and fruitful discussions.
%



\bibliographystyle{mnras}
\bibliography{references} 

\begin{thebibliography}{}
\makeatletter
\relax
\def\mn@urlcharsother{\let\do\@makeother \do\$\do\&\do\#\do\^\do\_\do\%\do\~}
\def\mn@doi{\begingroup\mn@urlcharsother \@ifnextchar [ {\mn@doi@}
  {\mn@doi@[]}}
\def\mn@doi@[#1]#2{\def\@tempa{#1}\ifx\@tempa\@empty \href
  {http://dx.doi.org/#2} {doi:#2}\else \href {http://dx.doi.org/#2} {#1}\fi
  \endgroup}
\def\mn@eprint#1#2{\mn@eprint@#1:#2::\@nil}
\def\mn@eprint@arXiv#1{\href {http://arxiv.org/abs/#1} {{\tt arXiv:#1}}}
\def\mn@eprint@dblp#1{\href {http://dblp.uni-trier.de/rec/bibtex/#1.xml}
  {dblp:#1}}
\def\mn@eprint@#1:#2:#3:#4\@nil{\def\@tempa {#1}\def\@tempb {#2}\def\@tempc
  {#3}\ifx \@tempc \@empty \let \@tempc \@tempb \let \@tempb \@tempa \fi \ifx
  \@tempb \@empty \def\@tempb {arXiv}\fi \@ifundefined
  {mn@eprint@\@tempb}{\@tempb:\@tempc}{\expandafter \expandafter \csname
  mn@eprint@\@tempb\endcsname \expandafter{\@tempc}}}

\bibitem[\protect\citeauthoryear{Acero et~al.,}{Acero
  et~al.}{2015}]{acero_fermi_2015}
Acero F.,  et~al., 2015, \mn@doi [ApJS] {10.1088/0067-0049/218/2/23}, 218, 23

\bibitem[\protect\citeauthoryear{Ackermann et~al.,}{Ackermann
  et~al.}{2015}]{ackermann_multiwavelength_2015}
Ackermann M.,  et~al., 2015, \mn@doi [ApJL] {10.1088/2041-8205/813/2/L41}, 813,
  L41

\bibitem[\protect\citeauthoryear{Ackermann et~al.,}{Ackermann
  et~al.}{2017}]{ackermann_gamma-ray_2017}
Ackermann M.,  et~al., 2017, \mn@doi [ApJL] {10.3847/2041-8213/aa5fff}, 837, L5

\bibitem[\protect\citeauthoryear{Ajello et~al.,}{Ajello
  et~al.}{2012}]{ajello_luminosity_2012}
Ajello M.,  et~al., 2012, \mn@doi [ApJ] {10.1088/0004-637X/751/2/108}, 751, 108

\bibitem[\protect\citeauthoryear{Ajello et~al.,}{Ajello
  et~al.}{2014}]{ajello_cosmic_2014}
Ajello M.,  et~al., 2014, \mn@doi [ApJ] {10.1088/0004-637X/780/1/73}, 780, 73

\bibitem[\protect\citeauthoryear{Arzoumanian et~al.,}{Arzoumanian
  et~al.}{2014}]{arzoumanian_gravitational_2014}
Arzoumanian Z.,  et~al., 2014, \mn@doi [ApJ] {10.1088/0004-637X/794/2/141},
  794, 141

\bibitem[\protect\citeauthoryear{Bansal, Taylor, Peck, Zavala  \&
  Romani}{Bansal et~al.}{2017}]{bansal_constraining_2017}
Bansal K.,  Taylor G.~B.,  Peck A.~B.,  Zavala R.~T.,   Romani R.~W.,  2017,
  \mn@doi [ApJ] {10.3847/1538-4357/aa74e1}, 843, 14

\bibitem[\protect\citeauthoryear{Barth, Ho  \& Sargent}{Barth
  et~al.}{2002}]{barth_stellar_2002}
Barth A.~J.,  Ho L.~C.,   Sargent W. L.~W.,  2002, \mn@doi [ApJL]
  {10.1086/339452}, 566, L13

\bibitem[\protect\citeauthoryear{Barth, Ho  \& Sargent}{Barth
  et~al.}{2003}]{barth_black_2003}
Barth A.~J.,  Ho L.~C.,   Sargent W. L.~W.,  2003, \mn@doi [ApJ]
  {10.1086/345083}, 583, 134

\bibitem[\protect\citeauthoryear{B\"{o}ttcher \& Dermer}{B\"{o}ttcher \&
  Dermer}{2002}]{bottcher_evolutionary_2002}
B\"{o}ttcher M.,  Dermer C.~D.,  2002, \mn@doi [ApJ] {10.1086/324134}, 564, 86

\bibitem[\protect\citeauthoryear{Caproni, Abraham, Motter  \& Monteiro}{Caproni
  et~al.}{2017}]{caproni_jet_2017}
Caproni A.,  Abraham Z.,  Motter J.~C.,   Monteiro H.,  2017, \mn@doi [ApJL]
  {10.3847/2041-8213/aa9fea}, 851, L39

\bibitem[\protect\citeauthoryear{Castignani, Haardt, Lapi, De~Zotti, Celotti
  \& Danese}{Castignani et~al.}{2013}]{castignani_black-hole_2013}
Castignani G.,  Haardt F.,  Lapi A.,  De~Zotti G.,  Celotti A.,   Danese L.,
  2013, \mn@doi [A\&A] {10.1051/0004-6361/201321424}, 560, A28

\bibitem[\protect\citeauthoryear{Charisi, Bartos, Haiman, Price-Whelan, Graham,
  Bellm, Laher  \& M\'{a}rka}{Charisi et~al.}{2016}]{charisi_population_2016}
Charisi M.,  Bartos I.,  Haiman Z.,  Price-Whelan A.~M.,  Graham M.~J.,  Bellm
  E.~C.,  Laher R.~R.,   M\'{a}rka S.,  2016, MNRAS, 463, 2145

\bibitem[\protect\citeauthoryear{Charisi, Haiman, Schiminovich  \&
  D'Orazio}{Charisi et~al.}{2018}]{charisi_testing_2018}
Charisi M.,  Haiman Z.,  Schiminovich D.,   D'Orazio D.~J.,  2018, \mn@doi
  [MNRAS] {10.1093/mnras/sty516}, 476, 4617

\bibitem[\protect\citeauthoryear{D'Orazio, Haiman  \& Schiminovich}{D'Orazio
  et~al.}{2015}]{dorazio_relativistic_2015}
D'Orazio D.~J.,  Haiman Z.,   Schiminovich D.,  2015, \mn@doi [Nature]
  {10.1038/nature15262}, 525, 351

\bibitem[\protect\citeauthoryear{De~Paolis, Ingrosso  \& Nucita}{De~Paolis
  et~al.}{2002}]{de_paolis_astrophysical_2002}
De~Paolis F.,  Ingrosso G.,   Nucita A.~A.,  2002, \mn@doi [A\&A]
  {10.1051/0004-6361:20020519}, 388, 470

\bibitem[\protect\citeauthoryear{Demorest et~al.,}{Demorest
  et~al.}{2013}]{demorest_limits_2013}
Demorest P.~B.,  et~al., 2013, \mn@doi [ApJ] {10.1088/0004-637X/762/2/94}, 762,
  94

\bibitem[\protect\citeauthoryear{Desvignes et~al.,}{Desvignes
  et~al.}{2016}]{desvignes_high-precision_2016}
Desvignes G.,  et~al., 2016, \mn@doi [MNRAS] {10.1093/mnras/stw483}, 458, 3341

\bibitem[\protect\citeauthoryear{Falomo, Kotilainen  \& Treves}{Falomo
  et~al.}{2002}]{falomo_black_2002}
Falomo R.,  Kotilainen J.~K.,   Treves A.,  2002, \mn@doi [ApJ]
  {10.1086/340642}, 569, L35

\bibitem[\protect\citeauthoryear{Ferrarese \& Merritt}{Ferrarese \&
  Merritt}{2000}]{ferrarese_fundamental_2000}
Ferrarese L.,  Merritt D.,  2000, \mn@doi [ApJL] {10.1086/312838}, 539, L9

\bibitem[\protect\citeauthoryear{{Foster} \& {Backer}}{{Foster} \&
  {Backer}}{1990}]{1990ApJ...361..300F}
{Foster} R.~S.,  {Backer} D.~C.,  1990, \mn@doi [\apj] {10.1086/169195}, \href
  {http://adsabs.harvard.edu/abs/1990ApJ...361..300F} {361, 300}

\bibitem[\protect\citeauthoryear{Gebhardt et~al.,}{Gebhardt
  et~al.}{2000}]{gebhardt_relationship_2000}
Gebhardt K.,  et~al., 2000, \mn@doi [ApJL] {10.1086/312840}, 539, L13

\bibitem[\protect\citeauthoryear{Ghisellini, Tavecchio, Foschini  \&
  Ghirlanda}{Ghisellini et~al.}{2011}]{ghisellini_transition_2011}
Ghisellini G.,  Tavecchio F.,  Foschini L.,   Ghirlanda G.,  2011, \mn@doi
  [MNRAS] {10.1111/j.1365-2966.2011.18578.x}, 414, 2674

\bibitem[\protect\citeauthoryear{Graham et~al.,}{Graham
  et~al.}{2015}]{graham_systematic_2015}
Graham M.~J.,  et~al., 2015, \mn@doi [MNRAS] {10.1093/mnras/stv1726}, 453, 1562

\bibitem[\protect\citeauthoryear{Inayoshi, Ichikawa  \& Haiman}{Inayoshi
  et~al.}{2018}]{inayoshi_gravitational_2018}
Inayoshi K.,  Ichikawa K.,   Haiman Z.,  2018, preprint, 1805, arXiv:1805.05334

\bibitem[\protect\citeauthoryear{Kelley, Blecha  \& Hernquist}{Kelley
  et~al.}{2017}]{kelley_massive_2017}
Kelley L.~Z.,  Blecha L.,   Hernquist L.,  2017, \mn@doi [MNRAS]
  {10.1093/mnras/stw2452}, 464, 3131

\bibitem[\protect\citeauthoryear{King et~al.,}{King
  et~al.}{2013}]{king_quasi-periodic_2013}
King O.~G.,  et~al., 2013, \mn@doi [MNRAS Letters] {10.1093/mnrasl/slt125},
  436, L114

\bibitem[\protect\citeauthoryear{Kormendy \& Ho}{Kormendy \&
  Ho}{2013}]{kormendy_coevolution_2013}
Kormendy J.,  Ho L.~C.,  2013, \mn@doi [ARA\&A]
  {10.1146/annurev-astro-082708-101811}, 51, 511

\bibitem[\protect\citeauthoryear{Liu, Shen, Bian, Loeb  \& Tremaine}{Liu
  et~al.}{2014}]{liu_constraining_2014}
Liu X.,  Shen Y.,  Bian F.,  Loeb A.,   Tremaine S.,  2014, \mn@doi [ApJ]
  {10.1088/0004-637X/789/2/140}, 789, 140

\bibitem[\protect\citeauthoryear{Liu et~al.,}{Liu
  et~al.}{2016}]{liu_systematic_2016}
Liu T.,  et~al., 2016, \mn@doi [ApJ] {10.3847/0004-637X/833/1/6}, 833, 6

\bibitem[\protect\citeauthoryear{Phinney}{Phinney}{2001}]{phinney_practical_2001}
Phinney E.~S.,  2001, arXiv:astro-ph/0108028

\bibitem[\protect\citeauthoryear{Plotkin, Markoff, Trager  \& Anderson}{Plotkin
  et~al.}{2011}]{plotkin_dynamical_2011}
Plotkin R.~M.,  Markoff S.,  Trager S.~C.,   Anderson S.~F.,  2011, \mn@doi
  [MNRAS] {10.1111/j.1365-2966.2010.18172.x}, 413, 805

\bibitem[\protect\citeauthoryear{Reardon et~al.,}{Reardon
  et~al.}{2016}]{reardon_timing_2016}
Reardon D.~J.,  et~al., 2016, \mn@doi [MNRAS] {10.1093/mnras/stv2395}, 455,
  1751

\bibitem[\protect\citeauthoryear{Rieger}{Rieger}{2007}]{rieger_supermassive_2007}
Rieger F.~M.,  2007, \mn@doi [Astrophys Space Sci] {10.1007/s10509-007-9467-y},
  309, 271

\bibitem[\protect\citeauthoryear{Rodriguez, Taylor, Zavala, Peck, Pollack  \&
  Romani}{Rodriguez et~al.}{2006}]{rodriguez_compact_2006}
Rodriguez C.,  Taylor G.~B.,  Zavala R.~T.,  Peck A.~B.,  Pollack L.~K.,
  Romani R.~W.,  2006, \mn@doi [ApJ] {10.1086/504825}, 646, 49

\bibitem[\protect\citeauthoryear{Romero, Chajet, Abraham  \& Fan}{Romero
  et~al.}{2000}]{romero_beaming_2000}
Romero G.~E.,  Chajet L.,  Abraham Z.,   Fan J.~H.,  2000, A\&A, 360, 57

\bibitem[\protect\citeauthoryear{Runnoe et~al.,}{Runnoe
  et~al.}{2015}]{runnoe_large_2015}
Runnoe J.~C.,  et~al., 2015, \mn@doi [ApJS] {10.1088/0067-0049/221/1/7}, 221, 7

\bibitem[\protect\citeauthoryear{Runnoe et~al.,}{Runnoe
  et~al.}{2017}]{runnoe_large_2017}
Runnoe J.~C.,  et~al., 2017, \mn@doi [MNRAS] {10.1093/mnras/stx452}, 468, 1683

\bibitem[\protect\citeauthoryear{Sandrinelli, Covino  \& Treves}{Sandrinelli
  et~al.}{2014}]{sandrinelli_quasi-periodicities_2014}
Sandrinelli A.,  Covino S.,   Treves A.,  2014, \mn@doi [ApJL]
  {10.1088/2041-8205/793/1/L1}, 793, L1

\bibitem[\protect\citeauthoryear{Sandrinelli, Covino, Dotti  \&
  Treves}{Sandrinelli et~al.}{2016}]{sandrinelli_quasi-periodicities_2016}
Sandrinelli A.,  Covino S.,  Dotti M.,   Treves A.,  2016, \mn@doi [AJ]
  {10.3847/0004-6256/151/3/54}, 151, 54

\bibitem[\protect\citeauthoryear{Sandrinelli et~al.,}{Sandrinelli
  et~al.}{2017}]{sandrinelli_gamma-ray_2017}
Sandrinelli A.,  et~al., 2017, \mn@doi [A\&A] {10.1051/0004-6361/201630288},
  600, A132

\bibitem[\protect\citeauthoryear{Sandrinelli, Covino, Treves, Holgado, Sesana,
  Lindfors  \& Ramazani}{Sandrinelli
  et~al.}{2018}]{sandrinelli_quasi-periodicities_2018}
Sandrinelli A.,  Covino S.,  Treves A.,  Holgado A.~M.,  Sesana A.,  Lindfors
  E.,   Ramazani V.~F.,  2018, \mn@doi [A\&A] {10.1051/0004-6361/201732550},
  615, A118

\bibitem[\protect\citeauthoryear{{Sesana}, {Vecchio}  \& {Colacino}}{{Sesana}
  et~al.}{2008}]{2008MNRAS.390..192S}
{Sesana} A.,  {Vecchio} A.,   {Colacino} C.~N.,  2008, \mn@doi [\mnras]
  {10.1111/j.1365-2966.2008.13682.x}, \href
  {http://adsabs.harvard.edu/abs/2008MNRAS.390..192S} {390, 192}

\bibitem[\protect\citeauthoryear{Sesana, Haiman, Kocsis  \& Kelley}{Sesana
  et~al.}{2018}]{sesana_testing_2018}
Sesana A.,  Haiman Z.,  Kocsis B.,   Kelley L.~Z.,  2018, \mn@doi [ApJ]
  {10.3847/1538-4357/aaad0f}, 856, 42

\bibitem[\protect\citeauthoryear{Shannon et~al.,}{Shannon
  et~al.}{2015}]{shannon_gravitational_2015}
Shannon R.~M.,  et~al., 2015, \mn@doi [Science] {10.1126/science.aab1910}, 349,
  1522

\bibitem[\protect\citeauthoryear{Shen, Greene, Strauss, Richards  \&
  Schneider}{Shen et~al.}{2008}]{shen_biases_2008}
Shen Y.,  Greene J.~E.,  Strauss M.~A.,  Richards G.~T.,   Schneider D.~P.,
  2008, \mn@doi [ApJ] {10.1086/587475}, 680, 169

\bibitem[\protect\citeauthoryear{Shen, Liu, Loeb  \& Tremaine}{Shen
  et~al.}{2013}]{shen_constraining_2013}
Shen Y.,  Liu X.,  Loeb A.,   Tremaine S.,  2013, \mn@doi [ApJ]
  {10.1088/0004-637X/775/1/49}, 775, 49

\bibitem[\protect\citeauthoryear{Springel et~al.,}{Springel
  et~al.}{2005}]{springel_simulations_2005}
Springel V.,  et~al., 2005, \mn@doi [Nature] {10.1038/nature03597}, 435, 629

\bibitem[\protect\citeauthoryear{Tavani, Cavaliere, Munar-Adrover  \&
  Argan}{Tavani et~al.}{2018}]{tavani_blazar_2018}
Tavani M.,  Cavaliere A.,  Munar-Adrover P.,   Argan A.,  2018, \mn@doi [ApJ]
  {10.3847/1538-4357/aaa3f4}, 854, 11

\bibitem[\protect\citeauthoryear{Tremaine et~al.,}{Tremaine
  et~al.}{2002}]{tremaine_slope_2002}
Tremaine S.,  et~al., 2002, \mn@doi [ApJ] {10.1086/341002}, 574, 740

\bibitem[\protect\citeauthoryear{Urry}{Urry}{2000}]{urry_blazars_2000}
Urry C.~M.,  2000, \mn@doi [AIP Conference Proceedings] {10.1063/1.1291728},
  522, 299

\bibitem[\protect\citeauthoryear{Urry \& Padovani}{Urry \&
  Padovani}{1995}]{urry_unified_1995}
Urry C.~M.,  Padovani P.,  1995, \mn@doi [PASP] {10.1086/133630}, 107, 803

\bibitem[\protect\citeauthoryear{Valtonen et~al.,}{Valtonen
  et~al.}{2008}]{valtonen_massive_2008}
Valtonen M.~J.,  et~al., 2008, \mn@doi [Nature] {10.1038/nature06896}, 452, 851

\bibitem[\protect\citeauthoryear{Verbiest et~al.,}{Verbiest
  et~al.}{2016}]{verbiest_international_2016}
Verbiest J. P.~W.,  et~al., 2016, \mn@doi [MNRAS] {10.1093/mnras/stw347}, 458,
  1267

\bibitem[\protect\citeauthoryear{Wang, Greene, Ju, Rafikov, Ruan  \&
  Schneider}{Wang et~al.}{2017}]{wang_searching_2017}
Wang L.,  Greene J.~E.,  Ju W.,  Rafikov R.~R.,  Ruan J.~J.,   Schneider D.~P.,
   2017, \mn@doi [ApJ] {10.3847/1538-4357/834/2/129}, 834, 129

\bibitem[\protect\citeauthoryear{Yan, Zhou, Zhang, Zhu  \& Wang}{Yan
  et~al.}{2018}]{yan_testing_2018}
Yan D.,  Zhou J.,  Zhang P.,  Zhu Q.,   Wang J.,  2018, arXiv:1804.05342
  [astro-ph]

\makeatother
\end{thebibliography}



\appendix


\bsp	
\label{lastpage}
\end{document}